# محاسبه‌ی ثابتهای الاستیک بلور Ce -γ

رضاییان ، پیمان [۱و۲]؛ جلالی اسد آبادی ، سعید [۱و۳]؛ امینی ، حامد [۱و۲]

[۱] گروه فیزیک دانشگاه اصفهان ، اصفهان
[۲] انجمن علمی دانشجویی گروه فیزیک دانشگاه اصفهان ، اصفهان
[۳] مرکز پژوهش دانش و فناوری نانو، دانشگاه اصفهان ، اصفهان

## چکیده

در این مقاله ثابتهای الاستیک سریم با استفاده از کد محاسباتی WIEN2k به روش امواج تخت بهبود یافته بعلاوه اوربیتال های موضعی با پتانسیل کامل (FP-APW+lo) بر اساس نظریه‌ی تابعی چگالی (DFT) و اعمال تقریب شیب تعمیم یافته (GGA) برای انرژی تبادلی – همبستگی محاسبه شده است . برای تعیین این ثابتها از مشتقات مرتبه دوم انرژی نسبت به پارامترهای شبکه با تبدیل ساختار اولیه (مکعب سطح مرکز دار) به ساختار تتراگونالی مرکزدار استفاده شده است .

## Calculation of Elastic Constants of γ -Ce

Rezaeian, Peiman[1,2] ; Jalali Asadabadi, Saeid[1,3] ; Amini, Hamed[1,2]

[1] Physics Department, Isfahan University, Isfahan
[2] Student Scientific society of physics Dept. of Isfahan University, Isfahan
[3] Research center for nano sciences and nano technology , The University of Isfahan, Isfahan

### Abstract

*In this paper we calculated the elastic constants of γ-Ce . The calculations were performed self-consistently using the full potential augmented plane wave plus local orbital (FP-APW+lo) method. We used the generalized gradient approximation (GGA) to calculate the exchange-correlation energy. The elastic constants were obtained from the second order derivatives of energy with respect to lattice parameters. In this work we introduced a method to impose a deformation to the primary structure to simplify our calculations changing an fcc structure to bct.*

PACS No. 62

## مقدمه

سریم اولین عنصر گروه لانتانیدها است. اوربیتال 4f عناصر این گروه در حال پر شدن می باشد و با توجه به ویژگیهای این نوع الکترونها ، بررسی خواص لانتانیدها از جمله ویژگیهای ساختاری آنها همواره مورد توجه فیزیکدانان ماده چگال بوده است [۱].

هرگاه به یک بلور تنش وارد شود، انرژی ذخیره شده در آن بر حسب ثابت های الاستیک و مولفه های تانسور کرنش برابر است با [۴]:

$$E = \frac{V_0}{2} \sum_{i,j=1}^{6} C_{ij}\varepsilon_i\varepsilon_j \qquad (1)$$

در این رابطه $C_{ij}$ ها مولفه های تانسور الاستیک هستند. این تانسور ، از مرتبه ٤ است که در کمترین تقارن بلورین ، ۲۱ مولفه مستقل دارد . در مورد یک تتراگونال تعداد مولفه های مستقل به ٦ و برای مکعب این تعداد به ۳ کاهش می یابد .

## نحوه محاسبات

برای به دست آوردن ثابتهای الاستیک باید انرژی را به صورت تابعی از پارامتر (های) شبکه بدانیم [۲] .

محاسبه برخی ثابتهای الاستیک ساختار مکعب سطح مرکز دار، مستلزم دانستن انرژی به صورت تابعی دو متغیره از پارامتر های شبکه است ؛ از آنجایی که به دست آوردن چنین تابعی کار دشواری است، نمی توان به صورت مستقیم از ساختار مکعب سطح مرکز دار استفاده کرد ؛ بنابراین باید ساختار های دیگری را در نظر بگیریم و با استفاده از روش تبدیل و محاسبه پارامتر های شبه جدید انرژی را به صورت یک تابع تک متغیره از آنها به دست آوریم.

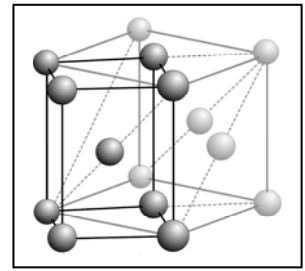

شکل ۱: تبدیل مکعب سطح مرکز دار به تتراگونال مرکزدار

## روش تبدیل

برای تغییر ساختار مکعب سطح مرکز دار، با وصل کردن اتم واقع در مرکز یکی از سطوح به دو اتم مجاور واقع در رئوس، ساختار تتراگونال مرکز دار ایجاد می شود. این کار معادل آن است که گفته شود دستگاه مکعب سطح مرکز دار را حول محور سوم به اندازه ۴۵ درجه دوران داده و به یک ساختار تتراگونال تبدیل می کنیم.

با این کار پارامتر های شبکه به این شکل تغییر می یابند :

$$a'_0 = b'_0 = \frac{a_0}{\sqrt{2}}$$
$$c'_0 = a_0 \qquad (2)$$

تحت این دوران مولفه های تانسور الاستیک تغییر می کند و با توجه به اینکه این تانسور از مرتبه ۴ می باشد اگر ماتریس دوران $\chi$ باشد آنگاه رابطه مولفه ها در دو دستگاه به صورت رابطه (۳) خواهد بود :

$$C'_{ijkl} = \sum \chi'_{il} \chi'_{jm} \chi'_{kn} \chi'_{lo} C_{lmno} \qquad (3)$$

در این روابط مولفه های پریم دار مربوط به ساختار تتراگونال و مولفه های بدون پریم مربوط به ساختار مکعبی سطح مرکز دار است. این رابطه را می توان به شکل زیر بازنویسی کرد :

$$C_{11} = C'_{33} \qquad (4)$$
$$C_{12} = C'_{33} - 2C'_{66}$$
$$C_{44} = C'_{11} - C'_{33} + C'_{66}$$

در این رابطه $C'_{11}$، $C'_{33}$ و $C'_{66}$ به صورت زیرمی باشند [۲].

$$C'_{33} = \frac{2c'_0}{a'^2_0} \frac{\partial^2 E}{\partial c'^2} \qquad (5-الف)$$

$$C'_{11} = \frac{2}{c'_0} \frac{\partial^2 E}{\partial a'^2} \qquad (5-ب)$$

$$C'_{66} = \frac{2}{c'_0 a'^2_0} \frac{\partial^2 E}{\partial \gamma'^2} \qquad (5-ج)$$

که در این روابط کمیتهای با اندیس صفر، پارامتر های شبکه در پایدارترین حالت بلور (کمینه منحنی انرژی) هستند. چنان که مشاهده می شود برای محاسبه این سه ثابت الاستیک، در مورد ساختار تتراگونال بر خلاف ساختار سطح مرکز دار نیازی نداریم انرژی را به صورت تابعی دو متغیره بدانیم.

برای بدست آوردن تابع انرژی، منحنی تغییرات انرژی ناشی از تغییر پارامتر های شبکه تتراگونال مرکز دار را رسم می کنیم. بدین ترتیب که هر پارامتر را پنج بار تغییر می دهیم و انرژی متناظر با آن را ثبت می کنیم، این نقاط را در مختصاتی که محور افقی آن پارامتر شبکه مورد نظر و محور عمودی آن انرژی است نشان می دهیم و با برازش یک منحنی درجه سه [۲،۳] شکل تابع انرژی را

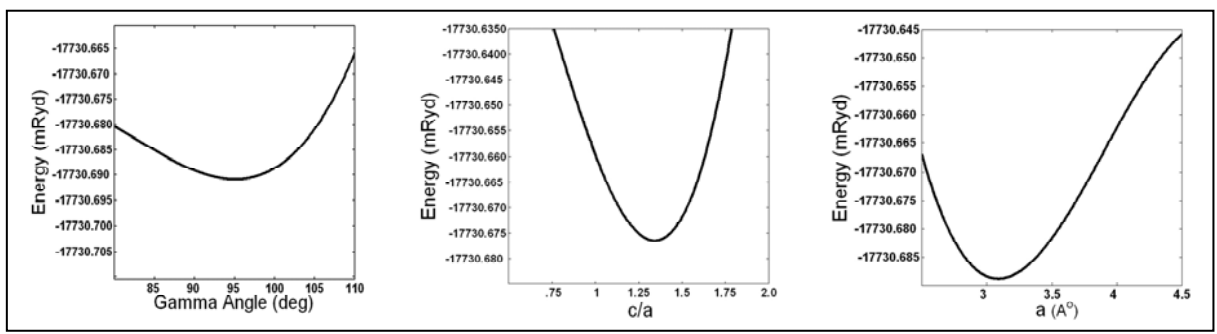

نمودار شماره ۱ : تغییرات انرژی بر حسب تغییرات پارامتر شبکه

$C'_{11} = 0.829\, Mbar$,

$C'_{33} = 0.537\, Mbar$

$C'_{66} = 6.8\, bar$

حال با استفاده از رابطه (4) می توان ثابتهای الاستیک بلور سریم را با ساختار مکعب سطح مرکز دار به شرح زیر محاسبه کرد :

$C_{11} = 0.538\, Mbar$

$C_{12} = 0.537\, Mbar$

$C_{44} = 0.291\, Mbar$

با مشخص بودن این ثابتها می توان اطلاعات مهمی در مورد ویژگی های ساختار بلور سریم از جمله مدول حجمی و سرعت صوت را در جهات مختلف این بلور بدست آورد[4] .

## سپاسگزاری

این کار با پشتیبانی مالی حوزه معاونت پژوهشی دانشگاه اصفهان از طریق طرح پژوهشی به شماره 821235 انجام شده است .

## مراجع

[1] Anna Delin, Lars Fast, and Börje Johansson; "Cohesive properties of the lanthanides: Effect of generalized gradient corrections and crystal structure", *Phys. Rev. B* **58**, 4345–4351 (1998)

[2] F. Jona , P.M. Marcus "Structural properties of copper", *Phys. Rev. B* **63**, 094113 (2001)

[3] S. L. Qiu. , P. M. Marcus "Elasticity of hcp nonmagnetic Fe under pressure", *Phys. Rev. B* **68**, 054103 (2003)

[4] C. Kittel "*Introduction to solid state physics*", 7th edition, John Wiley &Sons , Inc.(1996)

لازم به ذکر است که در استفاده از رابطه (5- الف) برای به دست آوردن $C'_{66}$ نیاز به تغییر زاویه $\gamma$ داریم . با تغییر این زاویه ($\gamma \neq 90$) ساختار به یک مونوکیلینیک مرکز دار تبدیل می شود که با توجه به عدم وجود چنین گروه فضایی در طبیعت با مشکل مواجه خواهیم شد . برای رفع این مشکل می توان تتراگونال مرکز دار را به یک تری کلینیک ساده تبدیل نمود و تغییر زاویه را روی آن اعمال کرد (شکل شماره 2).

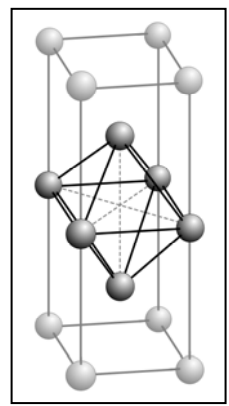

شکل 2: تبدیل تتراگونال مرکز دار به تری کلینیک ساده

## نتیجه گیری

در این مقاله با ارایه روش تبدیل ، ساختار مکعب سطح مرکز دار را به یک ساختار تتراگونال مرکز دار تبدیل کردیم و ثابتهای الاستیک تتراگونال مرکز دار را با بهره گیری از روابط (5) به دست آوردیم .

همان طور که قبلا ذکر شد بدین منظور باید پارامتر های شبکه را در پایدارترین حالت بلور بدانیم . این پارامتر ها را با به دست آوردن کمینه منحنی انرژی نمودار 1 محاسبه کرده ایم که عبارتند از $\gamma = 93°$، $c/a = 1.35$ و $a = 3.10 A$ و انحراف کمی از مقادیر تجربی $(c/a)_{real} = 1.4$، $\gamma_{real} = 90°$ و $a_{real} = 3.64 A$ نشان می دهند .

بدین ترتیب ثابتهای الاستیک تتراگونال برابر است با :